\documentclass[12pt]{article}
\usepackage{epsfig,latexsym,amssymb}
\begin{document}

\begin{center}
{\large \bf  ISCO, Lyapunov exponent and Kolmogorov-Sinai entropy for Kerr-Newman Black hole}
\end{center}

\vskip 5mm

\begin{center}
{\Large{Partha Pratim Pradhan\footnote{E-mail: pppradhan77@gmail.com,~ pppradhan@vsm.org.in}}}
\end{center}

\vskip  0.5 cm

{\centerline{\it Department of Physics}}
{\centerline{\it Vivekananda Satabarshiki Mahavidyalaya}}
{\centerline{\it Manikpara, Paschim Medinipur}}
{\centerline{\it West Bengal~721513, India}}

\vskip 1cm

\begin{abstract}

We compute the principal Lyapunov exponent and \emph{Kolmogorov-Sinai(KS) entropy} for Kerr-Newman black hole 
space-times and investigate the stability and instability of the equatorial circular geodesics via these exponents. 
We also show that the principal Lyapunov exponent and KS entropy can be expressed \emph{in terms of the radial equation 
of ISCO}(innermost stable circular orbit) for timelike circular geodesics. The other aspect we have studied that among the 
all possible circular geodesics, which encircle the central black-hole, the timelike circular geodesics  has 
the \emph{longest} orbital period i.e. $T_{timelike}>T_{photon}$, than the null circular geodesics (photon sphere) as 
measured by  asymptotic observers. Thus, the timelike circular geodesics provide the \emph{slowest way} to circle 
the Kerr-Newman black-hole. In fact, any stable timelike circular geodesics other than the ISCO traverses more slowly than 
the null circular geodesics.

\vspace{5mm} PACS number(s):  04.70.Bw, 05.50.+h

\end{abstract}

\section{Introduction}

Geodesic properties of blackholes have been examined for many years \cite{pp,pp1,pp2} to probe the stability or instability 
of circular geodesics using the effective potential. From the best of my knowledge, we know there were no attempt to made a link 
between non-linear Einstein's general theory of relativity and nonlinear dynamics. Particularly, Lyapunov exponent and KS entropy 
are two of them. In this article, we shall focus  on analytical calculations involving Lyapunov exponent and KS entropy in terms of 
the radial equation of circular geodesics around a black-hole space-time. This equatorial circular geodesics around a black-hole 
play an important role in general relativity for classification of the orbits. It also determines the important characteristics of 
the space-times and gives important information on the back ground geometry.

The author in \cite{car} computed the Lyapunov exponent to probe the instability of circular null geodesics in terms of the 
quasi-normal modes(QNMs) for spherically symmetric  space-times, but the focus there is on null circular geodesics.
It has been shown in this reference that by computing the Lyapunov exponent, which is the inverse of the instability time scale 
associated with the geodesic motion and in the eikonal limit QNMs of blackholes is determined by the parameters of the circular 
null geodesics.

Note however that the principal Lyapunov exponents($\lambda$) have been computed in \cite{car,clv} using a \emph{coordinate} 
time $t$,  where $t$ is measured by the asymptotic observers. Thus, these exponents are explicitly coordinate dependent and 
therefore have a degree of unphysicality. Here we compute the principal Lyapunov exponent($\lambda$) and Kolmogorov-Sinai 
entropy ($h_{ks}$) analytically by using the \emph{proper time} which is coordinate invariant.  Using Lyapunov exponent we
investigate the stability and instability of equatorial circular geodesics for Kerr-Newman black-hole space-times. Another 
interesting point we have studied here is that Lyapunov exponent  and KS entropy may be expressed  
\emph{in terms of the radial equation of ISCO}.

The author of \cite{hod} observed that the null circular geodesics is characterized by the
shortest possible orbital period as measured by the asymptotic observers among the all possible circular geodesics which encircle 
the central Kerr black-hole, thus  null circular geodesics provide the fastest way to circle black-hole. Here we would like 
to add that among the all possible circular geodesics, which encircle the central Kerr-Newman black-hole, the  timelike circular 
geodesics (ISCO) has the \emph{longest} orbital period than the null circular geodesics as measured by the asymptotic observers. 
Therefore, the timelike circular geodesics provide the \emph{slowest way} to circle the black-hole.

The plan of the Letter is as follows: in section \ref{def} we give the basic definition of Lyapunov exponent and 
also show that it may be expressed  in terms of the radial effective potential. In section \ref{ks} we provide the relation 
between Lyapunov exponent and KS entropy. In section \ref{cerp} we describe reciprocal of Critical exponent can be expressed
 in terms of the effective  radial potential.  In section \ref{ecgkn} we fully describe the equatorial circular geodesics, 
both time-like and null case for Kerr-Newman spacetimes. In section \ref{leisco} the Lyapunov
exponent can be expressed in terms of the ISCO equation and studied stability of time-like circular geodesics. In section 
\ref{ceeisco} we show that reciprocal of Critical exponent can also be expressed  in terms of the ISCO equation and 
we finally conclude with discussions in section \ref{dis}.

\section{\label{def}Lyapunov exponent and Radial potential:}

In any classical phase space the Lyapunov exponent gives a measure of the average rates of expansion and contraction of a trajectories
surrounding it.  They are the key indicators of chaos in any dynamical systems and also they are the asymptotic quantities defined 
locally in state space, and describe the exponential rate at which a perturbation to a trajectory of a system grows or decays with
time at a certain location in the state space.  
A positive Lyapunov exponent indicates a divergence between two nearby geodesics, i.e. the paths of such a system are extremely 
sensitive to changes of the initial conditions. A negative Lyapunov exponent implies a convergence between two nearby geodesics.
Lyapunov exponents can also distinguish among fixed points, periodic motions, quasi-periodic motions, and chaotic motions.

In classical physics,  an n-dimensional autonomous smooth dynamical system is governed by the differential equation\cite{naf} of 
the form

\begin{eqnarray}
\frac{dx}{dt} &=& F(x;M) ~.\label{dxdt}
\end{eqnarray}
where $t$ is defined usually as time parameter. Following \cite{osl, mott}, chaos may be quantified in terms of Lyapunov exponents
when the following prescriptions are maintained: a) the system is autonomous; b) the relevant part of the phase space is bounded; 
c) the invariant measure is normalizable; d) the domain of the time parameter is infinite. This definition signifies that the 
Lyapunov exponent is invariant under space diffeomorphisms of the form $u=\psi(x)$. As a result, chaos is a property of the physical
system and does not depend on the coordinates used to describe the system.

In general relativity(GR), there is no concept of absolute time, therefore the time parameter
forces us to consider equation (\ref{dxdt}) under spacetime diffeomorphism: $u=\psi(x)$, $d\tau=\eta(x) dt$. Thus the classical 
indicators of chaos like Lyapunov exponent and KS entropy explicitly depend on the choice of the time parameter. This 
noninvariant characterization implies that chaos is a property of the coordinate system rather than a property of the 
physical system.

Motivated by the work of Motter\cite{mott}, we find that chaos, as characterized by positive Lyapunov
exponents and positive Kolmogorov -Sinai entropy. They are coordinate invariants and transform according to

\begin{eqnarray}
\lambda_{i}^{\tau} &=& \frac{\lambda_{i}^{t}}{<\eta>_{t}} ~.\label{taut}
\end{eqnarray}

and

\begin{eqnarray}
h_{ks}^{\tau} &=& \frac{h_{ks}^{t}}{<\eta>_{t}} ~.\label{hks}
\end{eqnarray}

where $0< (<\eta>_{t})<\infty$ is the time average of $\eta=\frac{d\tau}{dt}$ over typical
trajectory and $i=1,...,n$, $n$ is the phase-space dimension. Transformation like
$u=\psi(x)$, $d\tau=\eta (x) dt$ is composed of a time re-parametrization followed by a space diffeomorphism. It is well known
that the Lyapunov exponents and Kolmogorov-Sinai Entropy are invariant under space diffeomorphism\cite{ott}.
In our previous work \cite{pp}, we have analysed in detail the derivation of Lyapunov exponents using proper time. Following this
the Lyapunov exponent may be expressed in terms of the radial potential

\begin{eqnarray}
\lambda &=&\pm \sqrt{\frac{(\dot{r}^{2})''}{2}} ~.\label{pot}
\end{eqnarray}
where we may defined $\dot{r}^2$ as radial potential or effective radial potential. In general the Lyapunov exponent 
come in $\pm$ pairs to conserve the volume of phase space.
The circular orbit is unstable  when the $\lambda$ is real, the circular orbit is stable 
when the $\lambda$ is imaginary and the circular orbit is marginally stable when $\lambda=0$.

\section{\label{ks}Kolmogorov-Sinai entropy and Lyapunove exponent:}

An important quantity which is related to the Lyapunov exponents is so called Kolmogorov-Sinai \cite{ks} entropy $(h_{ks})$,
gives a measure of the amount of information lost or gained by a chaotic orbit as it evolves. Alternatively it determines how a 
system is chaotic or disorder when $h_{ks}>0$ and non-chaotic for $h_{ks}=0$ \cite{ott}.

Following Pesin \cite{pes} it is equal to the sum of the positive Lyapunov exponents i.e

\begin{eqnarray}
h_{ks} &=& \sum_{\lambda_{i}>0}\lambda_{i} ~.\label{hk}
\end{eqnarray}
In 2-dimensional phase-space, there are two Lyapunov exponent, since $h_{ks}$ is equal
to the sum of positive Lyapunov exponent, therefore here the Kolmogorov-Sinai entropy
in terms of effective radial potential is given by
\begin{eqnarray}
h_{ks} &=& \sqrt{\frac{(\dot{r}^{2})''}{2}}~.\label{hkpot}
\end{eqnarray}

This entropy have played a crucial role in dynamical system  to check whether a trajectory  
is in disorder or not when it evolves with time. It is some sense different from the physical or statistical entropy, for example  
the entropy of the 2nd law of thermodynamics or blackhole entropy. Formally it is defined somewhat like entropy in statistical 
mechanics i.e it involves a partition of phase space.

\section{\label{cerp}Critical exponent and Radial potential:}

Following Pretorius and Khurana\cite{pk}, we can define  Critical exponent which is
the ratio of Lyapunov time scale $T_{\lambda}$ and Orbital time scale $T_{\Omega}$ may
be written as

\begin{eqnarray}
\gamma = \frac{\Omega}{2\pi\lambda}
=\frac{T_{\lambda}}{T_{\Omega}}=\frac{Lyapunov \, Time scale}{Orbital \, Time scale}~.\label{ce}
\end{eqnarray}
where we have introduced $T_{\lambda}=\frac{1}{\lambda}$ and $T_{\omega}=\frac{2\pi}{\Omega}$,
which is important for black-hole merger in the ring down radiation.
In terms of the square of the proper radial velocity ($\dot{r}^2$), Critical exponent can be
written as
\begin{eqnarray}
\gamma&=& \frac{T_{\lambda}}{T_{\Omega}}=
\frac{1}{2\pi}\sqrt{\frac{2\Omega ^2}{(\dot{r}^{2})''}} ~.\label{cerd}
\end{eqnarray}

Alternatively the reciprocal of critical exponent is proportional to the effective radial potential which is given by
\begin{eqnarray}
\frac{1}{\gamma}&=&\frac{T_{\Omega}}{T_{\lambda}}=
2\pi\sqrt{\frac{(\dot{r}^{2})''}{2\Omega^2}} ~.\label{ceinv}
\end{eqnarray}

\section{\label{ecgkn}Equatorial circular geodesics of the Kerr-Newman black-hole:}

The Einstein-Maxwell field of a stationary, axisymmetric, charged spinning body with mass $M$, charge $Q$ and angular
momentum parameter $a$ is described by the Kerr-Newman(KN) metric which in Boyer-Lindquist coordinates may be written as

\begin{eqnarray}
ds^2=-\frac{\Delta}{\rho^2}\, \left[dt-a\sin^2\theta\, d\phi \right]^2+
\frac{\sin^2\theta}{\rho^2}\,\left[(r^2+a^2)\,d\phi-a dt\right]^2
+\rho^2\left[\frac{dr^2}{\Delta}+(d\theta)^2\right] ~.\label{nkn}
\end{eqnarray}
where
\begin{eqnarray}
a&\equiv&\frac{J}{M}\nonumber\\
\rho^2 &\equiv& r^2+a^2\cos^2\theta \nonumber\\
\Delta &\equiv& r^2-2Mr+a^2+Q^2\equiv(r-r_{+})(r-r_{-}) ~.\label{nKn}
\end{eqnarray}
The metric is identical to the Schwarzschild black hole when $a=0,~Q=0$, Reissner Nordstr\"om black-hole when $a=0$ and 
Kerr black-hole when $Q=0$. The horizon occurs at $g_{rr}=\infty$ or $\Delta=0$ i.e.
\begin{eqnarray}
r_{\pm}&=&M\pm\sqrt{M^2-a^2-Q^2} \nonumber\\
\end{eqnarray}
The outer horizon $r_{+}$ is called event horizon and the inner horizon $r_{-}$ is called Cauchy horizon.

To compute the geodesics in the equatorial plane for the Kerr-Newman space-time we follow \cite{sch}. To determine the
geodesic motions of a test particle in this plane we set $\dot{\theta}=0$ and $\theta=constant=\frac{\pi}{2}$.

Therefore the necessary Lagrangian for this motion is given by
\begin{eqnarray}
{\cal L} =\frac{1}{2}\left[-\left(1-\frac{2M}{r}+\frac{Q^2}{r^2}\right)\,{\dot{t}}^2
-\left(\frac{4aM}{r}-\frac{2aQ^2}{r^2}\right)\,\dot{t}\,\dot{\phi}
+\frac{r^2}{\Delta}\,{\dot{r}}^2+\left(r^2+a^2+\frac{2Ma^2}{r}-\frac{a^2Q^2}{r^2}\right)
\,{\dot{\phi}}^2\right] ~.\label{lag}
\end{eqnarray}
The generalized momenta can be derived from it are
\begin{eqnarray}
p_{t} &=&-\left(1-\frac{2M}{r}+\frac{Q^2}{r^2}\right)\,\,\dot{t}-
  \left(\frac{2aM}{r}-\frac{aQ^2}{r^2}\right)\,\dot{\phi}=-E =Const ~.\label{pt}\\
p_{\phi} &=& -\left(\frac{2aM}{r}-\frac{aQ^2}{r^2}\right)\,\dot{t}
+\left[r^2+a^2+\frac{2Ma^2}{r}-\frac{a^2Q^2}{r^2}\right]\,\dot{\phi}=L=Const ~.\label{pphi}\\
p_{r} &=& \frac{r^2}{\Delta}\, \dot{r}  ~.\label{pr}
\end{eqnarray}
Here $(\dot{t},~\dot{r},~\dot{\phi})$ denotes differentiation with respect to proper time($\tau$). Since the Lagrangian 
does not depends on `t' and `$\phi$', so $p_{t}$ and $p_{\phi}$ are conserved quantities. The independence of the Lagrangian 
on `t' and `$\phi$' manifested, the stationarity and the axisymmetric character of the Kerr-Newman space-time.
The Hamiltonian is given by ${\cal H} = p_{t}\,\dot{t}+p_{\phi}\,\dot{\phi}+p_{r}\,\dot{r}-\cal L$.

In terms of the metric the Hamiltonian is
\begin{eqnarray}
2\cal H &=& -\left(1-\frac{2M}{r}+\frac{Q^2}{r^2}\right)\,\dot{t}^{2} -
  \left(\frac{4aM}{r}-\frac{2aQ^2}{r^2}\right)\dot{t}\,\dot{\phi}\nonumber \\[4mm] &&
  +\frac{r^2}{\Delta}\dot{r}^2+
  \left[r^2+a^2+\frac{2Ma^2}{r}-\frac{a^2Q^2}{r^2}\right]\dot{\phi}^2  ~.\label{hamh}
\end{eqnarray}
Since the Hamiltonian is independent of `t', therefore we can write it as

\begin{eqnarray}
2\cal H &=& -\left[\left(1-\frac{2M}{r}+\frac{Q^2}{r^2}\right)\,\dot{t}+
\left(\frac{2aM}{r}-\frac{aQ^2}{r^2}\right)\,\dot{\phi} \right]\,\dot{t}+\frac{r^2}{\Delta}\,\dot{r}^2
  +\nonumber \\[4mm] && \left[-\left(\frac{2aM}{r}-\frac{aQ^2}{r^2}\right)\,\dot{t}+\left(r^2+a^2+\frac{2Ma^2}{r}-
  \frac{a^2Q^2}{r^2}\right)\,\dot{\phi} \right]\,\dot{\phi} ~.\label{2ham}\\
        &=&-E\,\dot{t}+L\,\dot{\phi}+\frac{r^2}{\Delta}\,\dot{r}^2=\epsilon=const ~.\label{2hd}
\end{eqnarray}
Here $\epsilon=-1$ for time-like geodesics, $\epsilon=0$ for light-like geodesics and $\epsilon=+1$ for spacelike geodesics.
Solving equations (\ref{pt}) and (\ref{pphi}) for $\dot{\phi}$ and $\dot{t}$, we find
\begin{eqnarray}
\dot{\phi}&=&\frac{1}{\Delta}\left[\left(1-\frac{2M}{r}+\frac{Q^2}{r^2}\right)L
       +\left(\frac{2aM}{r}-\frac{aQ^2}{r^2}\right)E\right] ~.\label{uphi}\\
\dot{t}&=&\frac{1}{\Delta}\left[\left(r^2+a^2+\frac{2Ma^2}{r}-\frac{a^2Q^2}{r^2}\right)E
       -\left(\frac{2aM}{r}-\frac{aQ^2}{r^2}\right)L\right] ~.\label{ut}
\end{eqnarray}
Inserting these solutions in equations (\ref{2hd}), we obtain the radial equation for Kerr-Newman  space-time  which is given by
\begin{eqnarray}
r^2\dot{r}^{2} &=& r^2E^2+\left(\frac{2M}{r}-\frac{Q^2}{r^2}\right)\left(aE-L\right)^2 +\left(a^2E^2-L^2\right)+\epsilon\Delta ~.\label{radial}
\end{eqnarray}

\subsection{Circular null geodesics}

For null geodesics $\epsilon=0$, the radial equation (\ref{radial}) becomes

\begin{eqnarray}
r^2\dot{r}^{2} &=& r^2E^2+\left(\frac{2M}{r}-\frac{Q^2}{r^2}\right)\left(aE-L\right)^2 +\left(a^2E^2-L^2\right) ~.\label{nul}
\end{eqnarray}
The equations finding the radius of $r_{c}$ of the unstable circular `photon orbit' at $E=E_{c}$ and $L=L_{c}$ are
\begin{eqnarray}
E_{c}^2r_{c}^{2}+\left(\frac{2M}{r_{c}}-\frac{Q^2}{r_{c}^2}\right)\left(aE_{c}-L_{c}\right)^2
+\left(a^2E_{c}^2-L_{c}^2\right)&=& 0 ~.\label{rc}\\
2r_{c}E_{c}^2+\left(-\frac{2M}{r_{c}^2}+\frac{2Q^2}{r_{c}^3}\right)\left(aE_{c}-L_{c}\right)^2
  &=& 0 ~.\label{rc1}
\end{eqnarray}
Now introducing the impact parameter $D_{c}=\frac{L_{c}}{E_{c}}$, the above equations may be
written as
\begin{eqnarray}
r_{c}^{2}+\left(\frac{2M}{r_{c}}-\frac{Q^2}{r_{c}^2}\right)\left(a-D_{c}\right)^2
+\left(a^2-D_{c}^2\right) &=& 0  ~.\label{dc}\\
r_{c}-\left(\frac{M}{r_{c}^2}-\frac{Q^2}{r_{c}^3}\right)\left(a-D_{c}\right)^2 &=& 0 ~.\label{dc1}
\end{eqnarray}
From equation (\ref{dc1}) we have
\begin{eqnarray}
D_{c}&=& a\mp \frac{r_{c}^2}{\sqrt{Mr_{c}-Q^2}} ~.\label{dc2}
\end{eqnarray}
The equation (\ref{dc}) is valid if and only if $\mid D_{c}\mid>a$. For counter rotating
orbit, we have $|D_{c}-a|=-(D_{c}-a)$, which corresponds to upper sign in the above equation and co-rotating 
$|D_{c}-a|=+(D_{c}-a)$, which corresponds to lower sign in the above equation. Inserting equation (\ref{dc2}) in (\ref{dc}) we 
find an equation for the radius of null circular orbit
\begin{eqnarray}
 r_{c}^2-3Mr_{c}\pm 2a \sqrt{Mr_{c}-Q^2}+2Q^2 &=& 0  ~.\label{rnul}
\end{eqnarray}
When $Q=0$ we recover the well known result\cite{sch}.
Another important relation can be derived using equations (\ref{dc}) and (\ref{dc2}) for null circular orbits are
\begin{eqnarray}
D_{c}^2&=& a^2+r_{c}^2\left(\frac{3Mr_{c}-2Q^2}{Mr_{c}-Q^2} \right)~.\label{dc3}
\end{eqnarray}
Now we will derive an important physical quantity associated with the null circular geodesics is the angular frequency 
measured by asymptotic observers which is denoted by $\Omega_{c}$
\begin{eqnarray}
\Omega_{c}=\frac{\left[\left(1-\frac{2M}{r_{c}}+\frac{Q^2}{r_{c}^2}\right)D_{c}
+\left(\frac{2M}{r_{c}}-\frac{Q^2}{r_{c}^2}\right)a\right]}{\left[\left(r_{c}^2+
a^2+\frac{2Ma^2}{r_{c}}-\frac{a^2Q^2}{r_{c}^2}\right)
-a\left(\frac{2M}{r_{c}}-\frac{Q^2}{r_{c}^2}\right)D_{c}\right]}=\frac{1}{D_{c}}
~.\label{omnul}
\end{eqnarray}
Using equations  (\ref{dc2}) and (\ref{dc}) we show that the angular frequency $\Omega_{c}$ of the circular null 
geodesics is inverse of the impact parameter $D_{c}$, which generalizes the result of Kerr case\cite{sch} to the
Kerr-Newman black-hole space-time.
It proves that this is a \emph{general feature} of any stationary space-time.

The equation (\ref{nul}) governing null geodesics in terms of impact parameter at $E=E_{c}$ and $L=L_{c}$ can be
reduces to

\begin{eqnarray}
\dot{u}^{2} &=& E_{c}^2u^4(D_{c}-a)^2(u-u_{c})^2\left[M(2u+u_{c})-Q^2(u+u_{c})^2\right]  ~.\label{ud}
\end{eqnarray}
where
\begin{eqnarray}
u &=& \frac{1}{r},\,\, u_{c} = \frac{1}{r_{c}} \nonumber \\
r_{c} &=& \frac{3M}{2}\left[\frac{D_{c}-a}{D_{c}+a}\pm
\sqrt{\left(\frac{D_{c}-a}{D_{c}+a}\right)^2-\frac{8}{9}\left(\frac{Q}{M}\right)^2
\left(\frac{D_{c}-a}{D_{c}+a}\right)}\right] \nonumber\\
r_{c} &=& \frac{3M}{2}\left(\frac{D_{c}-a}{D_{c}+a}\right)\Xi  \\
\Xi &=& \left[1\pm\sqrt{1-\frac{8}{9}\left(\frac{D_{c}+a}{D_{c}-a}\right)
\left(\frac{Q}{M}\right)^2} \right]  ~.\label{rc}
\end{eqnarray}
Integrating equation (\ref{ud}) gives
\begin{eqnarray}
\tau \left[E_{c}(D_{c}-a)\right]&=& \pm \int\frac{du}
{u^2(u-u_{c})\sqrt{M(2u+u_{c})-Q^2(u+u_{c})^2}} ~.\label{uint}
\end{eqnarray}
To manifested the orbit in the equatorial plane, we can combine it with
the equation
\begin{eqnarray}
\dot{\phi} &=& \frac{E_{c}u^2}{3M\Xi(a^2u^2+a^2Q^2-2Mu+1)u_{c}}
\left[3MD_{c}u_{c}\Xi+2u(D_{c}+a)(Q^2u-2M)\right]~.\label{phid}
\end{eqnarray}
(which follows directly from equation(\ref{uphi})) to obtain
\begin{eqnarray}
\frac{du}{d\phi} &=& \frac{2a^2(D_{c}+a)(u-u_{c})(u-u_{+})(u-u_{-})
\sqrt{M(2u+u_{c})-Q^2(u+u_{c})^2}}
{\left[3MD_{c}u_{c}\Xi+2u(D_{c}+a)(Q^2u-2M)\right]} ~.\label{dudphi}
\end{eqnarray}
In the limit $Q=0$,~ $\Xi=2$, $\frac{du}{d\phi}$ is identical to \cite{sch}.
Integrating yields to obtain the trajectory for $\phi$
\begin{eqnarray}
\phi &=& \pm \frac{1}{2a^2(D_{c}+a)}\int
\frac{\left[3MD_{c}u_{c}\Xi+2u(D_{c}+a)(Q^2u-2M)\right]}
{(u-u_{+})(u-u_{-})(u-u_{c})\sqrt{M(2u+u_{c})-Q^2(u+u_{c})^2}}du ~.\label{intphi}
\end{eqnarray}
where $u_{\pm}=\frac{1}{r_{\pm}}$. The integral on the right hand side of equation (\ref{intphi}) is  contains partial fractions.

\subsection{Circular time-like geodesics}

For circular time-like geodesics equation (\ref{radial}) can be written as by
setting $\epsilon=-1$
\begin{eqnarray}
r^2\dot{r}^2 &=& r^2E^2+\left(\frac{2M}{r}-\frac{Q^2}{r^2}\right)\left(aE-L\right)^2
+\left(a^2E^2-L^2\right)-\Delta ~.\label{rtime}
\end{eqnarray}
where $E$ is the energy per unit mass of the particle describes the trajectory.

Now we shall find the radial equation of ISCO  that governing the time-like
circular geodesics in terms of reciprocal radius $u=1/r$ as the independent
variable, can be expressed as

\begin{eqnarray}
{\cal V} = u^{-4}\dot{u}^2=E^2+2Mu^3\left(aE-L\right)^2-u^4Q^2 \left(aE-L\right)^2 \nonumber\\
+\left(a^2E^2-L^2\right)u^2-(a^2+Q^2)u^2+2Mu-1 ~.\label{vu}
\end{eqnarray}

The conditions for the occurrence of circular orbits are at $r=r_{0}$ or reciprocal radius
$u=u_{0}$

\begin{eqnarray}
{\cal V} &=& 0 ~.\label{vu0}
\end{eqnarray}
and
\begin{eqnarray}
\frac{d{\cal V}}{du} &=& 0 ~.\label{dvdu}
\end{eqnarray}
Now setting $x=L_{0}-aE_{0}$, where $L_{0}$ and $E_{0}$ are the values of energy and angular momentum for circular orbits at the
radius $r_{0}=\frac{1}{u_{0}}$.
Therefore using (\ref{vu},~\ref{dvdu}) we get the following equations
\begin{eqnarray}
-x^2Q^2u_{0}^4+2Mx^2u_{0}^3-(x^2+2axE_{0})u_{0}^2-(a^2+Q^2)u_{0}^2+2Mu_{0}-1+E_{0}^2 &=& 0  ~.\label{x1}
\end{eqnarray}
and

\begin{eqnarray}
-2x^2Q^2u_{0}^3+3Mx^2u_{0}^2-(x^2+2axE_{0})u_{0}-(a^2+Q^2)u_{0}+M &=& 0  ~.\label{x2}
\end{eqnarray}
Using (\ref{x1},~\ref{x2}) we find an equation for $E_{0}^2$ as
\begin{eqnarray}
E_{0}^2 &=& 1-Mu_{0}+Mx^2u_{0}^3-x^2Q^2u_{0}^4  ~.\label{e2}
\end{eqnarray}
with the aid of equation (\ref{e2}), equation (\ref{x2}) gives us

\begin{eqnarray}
2axE_{0}u_{0} &=& x^2[3Mu_{0}^2-2Q^2u_{0}^3-u_{0}]-[(a^2+Q^2)u_{0}-M]  ~.\label{axe}
\end{eqnarray}
Eliminating $E_{0}$ between these equations, we have obtained the following
quadratic equation for $x^2$ i.e
\begin{eqnarray}
{\cal A}x^4+{\cal B}x^{2}+{\cal C} &=& 0   ~.\label{quad}
\end{eqnarray}
where

\begin{eqnarray}
{\cal A} &=& u_{0}^2 \left[[(3Mu_{0}-1)-2Q^2u_{0}^2]^2-4a^2(Mu_{0}^3-Q^2u_{0}^4) \right] \nonumber\\
{\cal B} &=& -2u_{0} \left[ (3Mu_{0}-1-2Q^2u_{0}^2)((a^2+Q^2)u_{0}-M)+2a^2u_{0}(1-Mu_{0})\right]
\nonumber\\
{\cal C} &=& \left[ (a^2+Q^2)u_{0}-M\right]^2   \nonumber
\end{eqnarray}
The solution of the equation (\ref{quad}) is
\begin{eqnarray}
x^2 &=& \frac{-{\cal B} \pm {\cal D}}{2 {\cal A}}   ~.\label{root}
\end{eqnarray}
where the discriminant of this equation is
\begin{eqnarray}
{\cal D} &=& 4au_{0} \, \Delta_{u_{0}} \sqrt{Mu_{0}-Q^2u_{0}^2}    ~.\label{discri}
\end{eqnarray}
and
\begin{eqnarray}
\Delta_{u_{0}} &=& (a^2+Q^2)u_{0}^2-2Mu_{0}+1   ~.\label{deltau}
\end{eqnarray}
The solution becomes simpler form by writing

\begin{eqnarray}
[(3Mu_{0}-1)-2Q^2u_{0}^2]^2-4a^2(Mu_{0}^3-Q^2u_{0}^4) &=& Z_{+}\,Z_{-}  ~.\label{z+}
\end{eqnarray}
where
\begin{eqnarray}
Z_{\pm} &=& (1-3Mu_{0}+2Q^2u_{0}^2) \pm 2a\sqrt{Mu_{0}^3-Q^2u_{0}^4}   ~.\label{z+-}
\end{eqnarray}
Thus we get the solution as
\begin{eqnarray}
x^2u_{0}^2 &=& \frac{-{\cal B} \pm {\cal D}}{Z_{+} Z_{-}}   ~.\label{x2u2}
\end{eqnarray}
Thus we find
\begin{eqnarray}
x^2u_{0}^2 &=& \frac{\Delta_{u}-Z_{\mp}}{Z_{\mp}}   ~.\label{zz}
\end{eqnarray}
Again we can write
\begin{eqnarray}
\Delta_{u_{0}}-Z_{\mp} &=& u_{0} \left[a\sqrt{u_{0}} \pm \sqrt{M-Q^2u_{0}} \right]^2  ~.\label{za}
\end{eqnarray}
Therefore the solution for $x$ thus may be written as
\begin{eqnarray}
x &=& - \frac{\left[a\sqrt{u_{0}} \pm \sqrt{M-Q^2u_{0}} \right]}{\sqrt{u_{0}{Z}_{\pm}}}  ~.\label{solx}
\end{eqnarray}
Here the upper sign in the foregoing equations applies to counter-rotating orbit, while the lower sign applies to 
co-rotating orbit.
Replacing the solution (\ref{solx}) for $x$ in equation (\ref{e2}), we obtain
the energy

\begin{eqnarray}
E_{0} &=&  \frac{1}{\sqrt{u_{0}{Z}_{\mp}}}\left[1-2Mu_{0} \mp au_{0}\sqrt{Mu_{0}-Q^2u_{0}^2} \right]  ~.\label{eng}
\end{eqnarray}
and the value of angular momentum associated with the circular orbit is given
by

\begin{eqnarray}
L_{0} &=& \mp \frac{1}{\sqrt{u_{0}{Z}_{\mp}}}\left[ \sqrt{M-Q^2u_{0}}\left( 1+a^2u_{0}^2\pm 2au_{0}\sqrt{Mu_{0}-Q^2u_{0}^2}\right) \pm aQ^2\sqrt{u_{0}^5}\right]  ~.\label{ang}
\end{eqnarray}

As we previously defined $E_{0}$ and $L_{0}$ followed by equations (\ref{eng}) and (\ref{ang}) are the energy and the angular
momentum per unit mass of a particle describing a circular orbit of radius $u_{0}$.
Therefore the minimum radius for a stable circular orbit will be obtained at a point of inflection of the function 
${\cal V}$ i.e we have to supply equations (\ref{vu0},~\ref{dvdu}) with the further equation

\begin{eqnarray}
\frac{d^{2}{\cal V}}{du^{2}}|_{u=u_{0}} &=& 0 ~.\label{d2vdu2}
\end{eqnarray}
Now we have to calculate

\begin{eqnarray}
\frac{d^{2}{\cal V}}{du^{2}} &=& \frac{1}{u}\left[ 6Mx^2u^2-8Q^2u^3x^2-2M\right] ~.\label{der2}
\end{eqnarray}
Using (\ref{zz}) we find
\begin{eqnarray}
\frac{d^{2}{\cal V}}{du^{2}}|_{u=u_{0}} &=& \frac{2}{u_{0}{Z}_{\mp}}\left[ \left(3M-4Q^2u_{0}\right) \Delta_{u_{0}}+ \left(4Q^2u_{0}-4M \right)Z_{\mp}  \right] ~.\label{derf}
\end{eqnarray}
Therefore the ISCO occurs at the reciprocal radius
\begin{eqnarray}
\frac{2}{u_{0}{Z}_{\mp}}\left[ \left(3M-4Q^2u_{0}\right) \Delta_{u_{0}}+ \left(4Q^2u_{0}-4M \right)Z_{\mp}  \right] &=& 0 ~.\label{isco}
\end{eqnarray}
or this can be written as

\begin{eqnarray}
\left(3Mu_{0}^2-4Q^2u_{0}^3\right)\left(a^2+Q^2 \right)+8Q^4u_{0}^3-12MQ^2u_{0}^2
\nonumber\\
\pm 8a\left(M-Q^2u_{0} \right)\sqrt{Mu_{0}^3-Q^2u_{0}^4}+6M^2u_{0}-M &=& 0 ~.\label{isco1}
\end{eqnarray}
Reverting to the variable $r_{0}$, we obtain the equation of ISCO for non-extremal Kerr-Newman black-hole  is given by
\begin{eqnarray}
Mr_{0}^3-6M^2r_{0}^2-3Ma^2r_{0}+9MQ^2r_{0} \mp 8a\left(Mr_{0}-Q^2\right)^{3/2}+4Q^2(a^2-Q^2) &=& 0 ~.\label{isco2}
\end{eqnarray}
Let $r_{0}=r_{ISCO}$  be the smallest real root of the equation, which will be the innermost stable circular orbit 
of the black-hole. Here $(-)$ sign indicates for direct orbit and
$(+)$ sign indicates for retrograde orbit.

\emph{Special cases:}
\begin{itemize}
\item  When $Q=0$, we recover the equation of ISCO for Kerr black-hole\cite{sch} which is given by
 \begin{eqnarray}
  r_{0}^2-6Mr_{0}\mp 8a\sqrt{Mr_{0}}-3a^2 &=& 0 ~.\label{iscoK}
 \end{eqnarray}
The smallest real root of this equation gives the radius of ISCO.

\item When $a=0$, we find the equation of ISCO for Reissner Nordstr{\o}m black hole\cite{sch} 
which is given by
 \begin{eqnarray}
  Mr_{0}^3-6M^2r_{0}^2+9MQ^2r_{0}-4Q^4 &=& 0 ~.\label{iscoRN}
 \end{eqnarray}
The radius of the ISCO can be obtained by finding the smallest real root of the above equation.

\item When $a=0,~Q=0$,  we get the radius of ISCO for Schwarzschild black hole is given by
\begin{eqnarray}
r_{0}-6M &=& 0 ~.\label{iscoSch}
\end{eqnarray}

\end{itemize}

For completeness, we include a treatment for trajectory of the time-like case.
Let $L_{0}$ and $E_{0}$ are the corresponding values of angular momentum and energy for circular orbit at the reciprocal 
radius $u=u_{0}$. Therefore
the equation (\ref{rtime}),  governing the time-like circular geodesics in terms of double root at $u=u_{0}$ can be reduces to

\begin{eqnarray}
u^{-4}\dot{u}^{2}=2M(L_{0}-a{E_{0}})^2(u-u_{0})^2\left[u+2u_{0}-(u+u_{0})^2
\frac{Q^2}{2M}-\frac{{L_{0}}^{2}-a^{2}{E_{0}}^{2}+a^2+Q^2}{2M(L_{0}-a{E_{0}})^2}
\right]  ~.\label{uc}
\end{eqnarray}

For $L_{0}$ and $E_{0}$ given by equations (\ref{ang}) and (\ref{eng}) (for
$u=u_{0}$) we find

\begin{eqnarray}
\frac{{L_{0}}^{2}-a^{2}{E_{0}}^{2}+a^2+Q^2}{2M(L_{0}-a{E_{0}})^2}
&=& \frac{1+3a^2{u_{0}}^{2}-Q^2{u_{0}}^{2}\pm 4a{u_{0}}\sqrt{M{u_{0}}-Q^2{u_{0}}^{2}}}{2(a\sqrt{u_{0}}\pm \sqrt{M-Q^2u_{0}})^{2}} ~.\label{uc1}
\end{eqnarray}

and

\begin{eqnarray}
u_{0}-\frac{{L_{0}}^{2}-a^{2}{E_{0}}^{2}+a^2+Q^2}{2M(L_{0}-a{E_{0}})^2}
   &=& -\frac{\Delta_{u_{0}}}{2(a\sqrt{u_{0}}\pm \sqrt{M-Q^2u_{0}})^{2}} ~.\label{uc2}
\end{eqnarray}

Therefore the equation  (\ref{uc}) can be rewritten as
\begin{eqnarray}
\dot{u}^{2} &=& 2Mu^{4}(L_{0}-a{E_{0}})^2(u-u_{0})^2(-Q_{*}^{2}u^2+bu+c) ~.\label{udot}
\end{eqnarray}

or this can be written as

\begin{eqnarray}
\dot{u}^{2} &=& 2Mu^{4}{x_{0}}^2(u-u_{0})^2(-Q_{*}^{2}u^2+\alpha u+\beta) ~.\label{udot1}
\end{eqnarray}

where
\begin{eqnarray}
 Q_{*}^{2}&=&\frac{Q^2}{2M}\nonumber\\
 x_{0} &=& L_{0}-a{E_{0}} \nonumber\\
  \alpha &=& 1-2u_{0}Q_{*}^{2} \nonumber\\
  \beta &=& -u_{*}- Q_{*}^{2}{u_{0}}^2 \nonumber\\
  u_{*} &=& -u_{0}+\frac{\Delta_{u_{0}}}{2(a\sqrt{u_{0}}\pm \sqrt{M-Q^2u_{0}})^{2}} ~.\label{bc}
\end{eqnarray}
It can be easily seen that $u_{*}$ defines the reciprocal radius of the
orbit of the 2nd kind.

Therefore the appropriate solution of equation (\ref{udot1}) is
\begin{eqnarray}
\tau &=& \frac{1}{x_{0}\sqrt{2M}}\int \frac{du}{u^2(u-u_{0})\sqrt{-Q_{*}^{2}u^2+
\alpha u+\beta}} ~.\label{tau1}
\end{eqnarray}
Again using equation (\ref{udot1}) with
\begin{eqnarray}
\dot{\phi} &=& \frac{u^2}{\Delta_{u}} \left[L_{0}-(2Mu-Q^2u^2)x_{0}\right]
  ~.\label{phi1}
\end{eqnarray}
we find the trajectory
\begin{eqnarray}
\phi &=& \frac{1}{x_{0}(a^2+Q^2)\sqrt{2M}}\int
\frac{\left[L_{0}-(2Mu-Q^2u^2)x_{0}\right]}{(u-u_{+})(u-u_{-})(u-u_{0})
\sqrt{-Q_{*}^{2}u^2+\alpha u+\beta}}du ~.\label{phi2}
\end{eqnarray}

\subsubsection{Angular Velocity of Time-like Circular Orbit}

Now we compute the orbital angular velocity for time-like circular geodesics at $r=r_{0}$
is given by
\begin{eqnarray}
\Omega_{0}=\frac{\dot{\phi}}{\dot{t}}
      =\frac{\left[\left(1-\frac{2M}{r_{0}}+\frac{Q^2}{r_{0}^2}\right)L_{0}
       +\left(\frac{2M}{r_{0}}-\frac{Q^2}{r_{0}^2}\right)aE_{0}\right]}{\left[\left(r_{0}^2+a^2+
       \frac{2Ma^2}{r_{0}}-\frac{a^2Q^2}{r_{0}^2}\right)E_{0}
       -a\left(\frac{2M}{r_{0}}-\frac{Q^2}{r_{0}^2}\right)L_{0}\right]}~.\label{omet}
\end{eqnarray}
Again this can be rewritten as
\begin{eqnarray}
\Omega_{0}=\frac{\left[L_{0}-2Mu_{0}x+Q^2u_{0}^2x \right]u_{0}^2}{(1+a^2u_{0}^2)E_{0}-2aMu_{0}^3x+aQ^2u_{0}^4x}~.\label{omeu}
\end{eqnarray}
Now the previously mentioned expression can be simplified as
\begin{eqnarray}
L_{0}-2Mu_{0}x+Q^2u_{0}^2x &=& \mp \frac{\sqrt{M-Q^2u_{0}}}{\sqrt{u_{0}{Z_{\mp}}}}\Delta_{u_{0}}~.\label{os}\\
(1+a^2u_{0}^2)E_{0}-2aMu_{0}^3x+aQ^2u_{0}^4x &=&
\frac{\Delta_{u_{0}}}{Z_{\mp}}\left( 1\mp a\sqrt{Mu_{0}^3-Q^2u_{0}^4}\right)~.\label{os1}
\end{eqnarray}
Substituting (\ref{os}) and  (\ref{os1}) into  (\ref{omeu}) we get the angular velocity for circular time-like geodesics 
is given by
\begin{eqnarray}
\Omega_{0}=\mp \frac{\sqrt{Mu_{0}^3-Q^2u_{0}^4}}{1 \mp a\sqrt{Mu_{0}^3-Q^2u_{0}^4}}~.\label{omef}
\end{eqnarray}
Reverting to the variable $r_{0}$, we obtain the angular velocity for time like circular orbit
is
\begin{eqnarray}
\Omega_{0}=\mp \frac{\sqrt{Mr_{0}-Q^2}}{r_{0}^2 \mp a\sqrt{Mr_{0}-Q^2}}~.\label{omer}
\end{eqnarray}
Correspondingly the time period for time like circular orbit is
\begin{eqnarray}
T_{0}=\frac{2\pi}{\Omega}=\mp 2\pi \frac{r_{0}^2 \mp a\sqrt{Mr_{0}-Q^2}}{\sqrt{Mr_{0}-Q^2}}~.\label{timep}
\end{eqnarray}
In the limit $a=Q=0$, this equation verifies the relativistic Kepler's law $T_{0}^{2}\propto r_{0}^{3}$ for Schwarzschild black-hole.

The rotational velocity with respect to the locally non-rotating observers is given by
\begin{eqnarray}
v^{\phi}=\frac{\left[ \sqrt{Mu_{0}-Q^2u_{0}^2}\left( 1+a^2u_{0}^2\pm 2au_{0}\sqrt{Mu_{0}-Q^2u_{0}^2}\right) \pm aQ^2\sqrt{u_{0}^5}\right]}{\left[1 \mp a\sqrt{Mu_{0}^3-Q^2u_{0}^4}\right]\sqrt{\Delta_{u_{0}}}}~.\label{omef}
\end{eqnarray}
Here we may note that we recover the photon spheres equation (\ref{rnul}) for KN spacetimes by taking the limit 
$E_{0}\rightarrow \infty$, when
\begin{eqnarray}
Z_{\pm} = 1-3Mu_{0}+2Q^2u_{0}^2\pm 2a\sqrt{Mu_{0}^3-Q^2u_{0}^4}=0   ~.\label{z0}
\end{eqnarray}

or alternatively

\begin{eqnarray}
 r_{0}^2-3Mr_{0}\pm 2a \sqrt{Mr_{0}-Q^2}+2Q^2 &=& 0  ~.\label{rul}
\end{eqnarray}
The above equation describe the radius of circular photon sphere equation (\ref{rnul}) at $r_{0}=r_{c}$. Here $(-)$ sign 
indicates for direct orbit and $(+)$ sign indicates for retrograde orbit. The real positive root of the equation
is the closest circular photon orbit of the black-hole.

\subsubsection{Ratio of Angular velocity of time like circular orbit to null circular Orbit}

Since we have already proved that for time-like circular geodesics the angular velocity is given by from equation (\ref{omef})
\begin{eqnarray}
\Omega_{0}=\mp \frac{\sqrt{Mu_{0}^3-Q^2u_{0}^4}}{1 \mp a\sqrt{Mu_{0}^3-Q^2u_{0}^4}} \nonumber\\
\end{eqnarray}
Again we obtained for circular null geodesics $\Omega_{c}=\frac{1}{D_{c}}$, so we can deduce similar expression for it is given by
\begin{eqnarray}
\Omega_{c}=\mp \frac{\sqrt{Mu_{c}^3-Q^2u_{c}^4}}{1 \mp a\sqrt{Mu_{c}^3-Q^2u_{c}^4}} ~.\label{omec}
\end{eqnarray}
Resultantly we obtain the ratio of angular frequency for time-like circular  geodesics to the angular frequency for null circular
geodesics is
\begin{eqnarray}
\frac{\Omega_{0}}{\Omega_{c}}&=& \left(\frac{\sqrt{Mr_{0}-Q^2}}{\sqrt{Mr_{c}-Q^2}}\right)
\left(\frac{r_{c}^2\mp a\sqrt{Mr_{c}-Q^2}}{r_{0}^2\mp a\sqrt{Mr_{0}-Q^2}}\right) ~.\label{ratio}
\end{eqnarray}
For $r_{0}=r_{c}$, $\Omega_{0}=\Omega_{c}$,i.e , when the radius of time-like circular geodesics is equal to the radius of null 
circular geodesics, the angular frequency corresponds to that geodesic are equal, which demands that the intriguing physical 
phenomena could occur in the curved space-time, for example, possibility of exciting quasi normal modes(QNM) by orbiting
particles, possibly leading to instabilities of the curved space-time \cite{car}.

For $r_{0}>r_{c}$, we have already proved that for Schwarzschild black-hole  and Reissner Nordstr{\o}m black-hole \cite{pp}
the null circular geodesics have the largest angular frequency as measured by asymptotic observers than the time-like 
circular geodesics. We therefore conclude that null circular geodesics provide the fastest way to circle black holes \cite{hod}.
This generalizes the case of spherically symmetry Schwarzschild black-hole and Kerr Black-hole \cite{hod} to the more general case 
of stationary, axi-symmetry Kerr-Newman space-times.

Now the ratio of time period of time-like circular  geodesics to the  time period of
null circular geodesics is given by
\begin{eqnarray}
\frac{T_{0}}{T_{c}}&=& \left(\frac{\sqrt{Mr_{c}-Q^2}}{\sqrt{Mr_{0}-Q^2}}\right)
\left(\frac{r_{0}^2\mp a\sqrt{Mr_{0}-Q^2}}{r_{c}^2\mp a\sqrt{Mr_{c}-Q^2}}\right) ~.\label{ratio1}
\end{eqnarray}
This ratio is valid for $r_{0}\neq r_{c}$. For $r_{0}=r_{c}$, $T_{0}=T_{c}$, i.e. time period of both geodesics are equal, which
possibly leading to the excitations of QNM.
For $r_{0}>r_{c}$, $T_{0}>T_{c}$, which implies that the orbital period of time-like circular geodesics is greater than the
orbital period of null circular geodesics. For $r_{0}=r_{ISCO}$ and $r_{c}=r_{photon}$,  the ratio of time period 
for ISCO ($r_{0}=r_{ISCO}$) to the  time period for photon-sphere $(r_{c}=r_{photon})$ for Kerr-Newman black-hole is given by
\begin{eqnarray}
\frac{T_{ISCO}}{T_{photon}}&=& \left(\frac{\sqrt{Mr_{c}-Q^2}}{\sqrt{Mr_{0}-Q^2}}\right)
\left(\frac{r_{0}^2\mp a\sqrt{Mr_{0}-Q^2}}{r_{c}^2\mp a\sqrt{Mr_{c}-Q^2}}\right)~.\label{isphkn}
\end{eqnarray}
This implies that $T_{ISCO}>T_{photon}$, this imlies that the orbital period of time-like circular geodesics is larger than 
the null circular geodesics. Thus we conclude that timelike circular geodesics  (ISCO) provide the 
\emph{slowest way} to circle the Kerr-Newman black-hole among all circular geodesics.

\subsubsection{Marginally bound circular orbit}

When a particle at rest at infinity falling towards the black-hole, we call the situation is marginally bound circular 
orbit. Using equations (\ref{eng}) and
(\ref{solx}), the radius of the marginally bound circular orbit with $E_{0}^2=1$ is given by
\begin{eqnarray}
 1=x^2u_{0}^2=\frac{\left[au_{0}\pm \sqrt{Mu_{0}-Q^2u_{0}^2} \right]^2}{Z_{\mp}}  ~.\label{marg}
\end{eqnarray}
or
\begin{eqnarray}
Z_{\mp} =\frac{Mu_{0}-Q^2u_{0}^2}{Mu_{0}}\left[au_{0}\pm \sqrt{Mu_{0}-Q^2u_{0}^2} \right]^2  ~.\label{marg1}
\end{eqnarray}
or
\begin{eqnarray}
Mu\left[(1-3Mu_{0}+2Q^2u_{0}^2) \pm 2a\sqrt{Mu_{0}^3-Q^2u_{0}^4}\right]=(Mu_{0}-Q^2u_{0}^2)\left[au_{0}\pm \sqrt{Mu_{0}-Q^2u_{0}^2} \right]^2  ~.\label{marg2}
\end{eqnarray}
After simplification we obtain the following form for marginally bound circular orbit is given by
\begin{eqnarray}
(a^2-Q^2)Q^2u_{0}^3+M(4Q^2-a^2)u_{0}^2-4M^2u_{0}\mp (4aMu_{0}-2aQ^2u_{0}^2)\sqrt{Mu_{0}-Q^2u_{0}^2}+M=0 ~.\label{marg3}
\end{eqnarray}
In terms of $r_{0}$ we can written it as
\begin{eqnarray}
Mr_{0}^3-4M^2r_{0}^2-Ma^2r_{0}+4MQ^2r_{0} \mp (4aMr_{0}-2aQ^2)\sqrt{Mr_{0}-Q^2}+Q^2(a^2-Q^2)= 0 ~.\label{marg4}
\end{eqnarray}

Let $r_{0}=r_{mb}$ be the real smallest root of the above equation, which will be  the closest bound circular orbit to
the black-hole.

\emph{Special Cases:}

In the above equation if we take the following limits: 
\begin{itemize}
 \item When $Q=0$, we obtain the equation of marginally bound circular orbit for Kerr black-hole which is given by
\begin{eqnarray}
r_{0}^2-4Mr_{0} \mp 4a\sqrt{Mr_{0}}-a^2 &=& 0 ~.\label{mbK}
\end{eqnarray}
The smallest real root of this equation gives the marginally bound circular orbit to the black-hole.
\item When $a=0$, we find the equation of marginally bound circular orbit for Reissner
Nordstr{\o}m Black hole which is given by

\begin{eqnarray}
Mr_{0}^3-4M^2r_{0}^2+4MQ^2r_{0}-Q^4 &=& 0 ~.\label{mbRN}
\end{eqnarray}
The radius of the marginally bound circular orbit $r_{0}=r_{mb}$ can be obtained by finding the smallest real root of
the above equation.

\item When  $a=0,~Q=0$,  we get the radius of marginally bound circular orbit for Schwarzschild black hole
which is given by
\begin{eqnarray}
r_{0}-4M &=& 0 ~.\label{mbSch}
\end{eqnarray}

\end{itemize}

\section{\label{leisco}Lyapunov exponent and Equation of ISCO:}

Now we evaluate the  Lyapunov exponent and KS entropy in terms of the radial equation of ISCO  as follows, 
using equation (\ref{pot}) one obtains

\begin{eqnarray}
\lambda=h_{ks}=\sqrt{\frac{-\left(Mr_{0}^3-6M^2r_{0}^2-3Ma^2r_{0}+9MQ^2r_{0} \mp 8a\left(Mr_{0}-Q^2\right)^{3/2}+4Q^2(a^2-Q^2)\right)}
{r_{0}^4\left(r_{0}^2-3Mr_{0}\mp2a\sqrt{Mr_{0}-Q^2}+2Q^2\right)}} ~.\label{lyisco1}
\end{eqnarray}

Circular geodesic motion of the test particle to be exists when both energy (\ref{eng}) and angular momentum (\ref{ang}) are real 
and finite, therefore we must have $r_{0}^2-3Mr_{0}\mp2a\sqrt{Mr_{0}-Q^2}+2Q^2>0$ and $r_{0}>\frac{Q^{2}}{M}$.

It can be easily seen from the above relation that L.H.S is the expression for Lyapunov exponent and KS entropy and R.H.S is
the equation of ISCO for Kerr-Newman black-hole. So the time-like circular geodesics of Kerr-Newman black-hole are stable when

\begin{eqnarray}
Mr_{0}^3-6M^2r_{0}^2-3Ma^2r_{0}+9MQ^2r_{0} \mp 8a\left(Mr_{0}-Q^2\right)^{3/2}+4Q^2(a^2-Q^2)>0 ~.\label{stable}
\end{eqnarray}
such that $\lambda$ or $h_{KS}$ is imaginary, the circular geodesics are unstable when

\begin{eqnarray}
Mr_{0}^3-6M^2r_{0}^2-3Ma^2r_{0}+9MQ^2r_{0} \mp 8a\left(Mr_{0}-Q^2\right)^{3/2}+4Q^2(a^2-Q^2)<0 ~.\label{unstable}
\end{eqnarray}
i.e $\lambda$ or $h_{KS}$  is real and the time like circular geodesics is marginally stable when
\begin{eqnarray}
Mr_{0}^3-6M^2r_{0}^2-3Ma^2r_{0}+9MQ^2r_{0} \mp 8a\left(Mr_{0}-Q^2\right)^{3/2}+4Q^2(a^2-Q^2)=0 ~.\label{marg}
\end{eqnarray}
such that $\lambda$ or $h_{KS}$ is zero.

\emph{Special Cases:}

\begin{itemize}
 \item  For Kerr black hole $Q=0$, the Lyapunov exponent and KS entropy for timelike circular geodesics are
\begin{eqnarray}
\lambda_{Kerr}=h_{ks}=\sqrt{\frac{-M\left(r_{0}^2-6M^2r_{0} \mp 8a\sqrt{Mr_{0}} -3a^2\right)}
{r_{0}^3\left(r_{0}^2-3Mr_{0}\mp2a\sqrt{Mr_{0}}\right)}} ~.\label{lyker}
\end{eqnarray}
 
 \item For Reissner Nordstr{\o}m black-hole $a=0$, the Lyapunov exponent and KS entropy for timelike circular geodesics are
\begin{eqnarray}
\lambda_{RN}=h_{ks}= \sqrt{\frac{-(Mr_{0}^{3}-6M^{2}r_{0}^{2}+9MQ^{2}r_{0}-4Q^{4})}{r_{0}^{4}
(r_{0}^{2}-3Mr_{0}+2Q^{2})}}~.\label{marn}
\end{eqnarray}
 
 \item For Schwarzschild black hole $a=Q=0$, the Lyapunov exponent and KS entropy in terms of ISCO equation  are
\begin{eqnarray}
\lambda_{Sch}=h_{ks} =\sqrt{-\frac{M(r_{0}-6M)}{r_{0}^{3}(r_{0}-3M)}}~.\label{marg3}
\end{eqnarray}
 
\end{itemize}

\subsection{Lyapunov exponent and Null circular geodesics:}

For null circular geodesics the Lyapunov exponent and KS entropy are given by

\begin{eqnarray}
\lambda_{Null} = h_{ks}=\sqrt{\frac{(L_{c}-aE_{c})^2(3Mr_{c}-4Q^2)}{r_{c}^6}}
 ~.\label{len2}
\end{eqnarray}
Since $(L_{c}-aE_{c})^2 \geq 0$ and $r_{c}>\frac{4}{3}\frac{Q^2}{M}$, therefore $\lambda_{Null}$
is real so the null circular geodesics are unstable.

\emph{Special Cases:}

\begin{itemize}

 \item  For Kerr black hole $Q=0$, the Lyapunov exponent and KS entropy for null circular geodesics are
\begin{eqnarray}
\lambda_{Null} = h_{ks}= \sqrt{\frac{3M(L_{c}-aE_{c})^2}{r_{c}^{5}}}~.\label{len3}
\end{eqnarray}

\item For Reissner Nordstr{\o}m black-hole $a=0$, so the  Lyapunov exponent and KS entropy for null circular geodesics are
\begin{eqnarray}
\lambda_{Null} = h_{ks} =\sqrt{\frac{{L_{c}}^2(3Mr_{c}-4Q^2)}{r_{c}^{6}}} ~.\label{len4}
\end{eqnarray}
So the geodesics are unstable since $\lambda_{Null}$ is real for $r_{c}>\frac{4}{3}\frac{Q^2}{M}$.

\item For Schwarzschild black hole $a=Q=0$, the Lyapunov exponent and KS entropy  are
\begin{eqnarray}
\lambda_{Null} = h_{ks}=\sqrt{\frac{3M{L_{c}}^2}{r_{c}^{5}}}~.\label{len5}
\end{eqnarray}
It can be easily check that for $r_{c}=3M$, $\lambda_{Null}$ is real, so the
Schwarzschild photon sphere are unstable.

\end{itemize}

\section{\label{ceeisco}Critical exponent and Equation of ISCO:}

Now we compute the reciprocal of Critical exponent in terms of ISCO equation for Kerr-Newman blackhole to be, by using  
 equation (\ref{ceinv})

\begin{eqnarray}
\frac{1}{\gamma}=2\pi\frac{\sqrt{-( Mr_{0}^3-6M^2r_{0}^2-3Ma^2r_{0}+9MQ^2r_{0} \mp 8a\left(Mr_{0}-Q^2\right)^{3/2}+4Q^2(a^2-Q^2))
(r_{0}^2\mp a\sqrt{Mr_{0}-Q^2})^2}}{\sqrt{r_{0}^4(Mr_{0}-Q^2)(r_{0}^2-3Mr_{0}\mp2a\sqrt{Mr_{0}-Q^2}+2Q^2)}} ~.\label{ceisco}
\end{eqnarray}

\emph{Special Cases:}

\begin{itemize}

\item For Kerr black hole $Q=0$ , the reciprocal of Critical exponent in terms of ISCO equation  are
\begin{eqnarray}
\frac{1}{\gamma} &=& 2\pi \frac{\sqrt{-\left[r_{0}^2-6Mr_{0}\mp 8a\sqrt{Mr_{0}}-3a^2\right](r_{0}\sqrt{r_{0}}\mp a\sqrt{M})^2}}
{\sqrt{r_{0}^3(r_{0}^2-3Mr_{0}\mp2a\sqrt{Mr_{0}})}}  ~.\label{marg1}
\end{eqnarray}

\item For Reissner Nordstr{\o}m black-hole $a=0$, the reciprocal of Critical exponent in terms of ISCO equation  are
\begin{eqnarray}
\frac{1}{\gamma}&=& \frac{\sqrt{-\left[Mr_{0}^3-6M^2r_{0}^2+9MQ^2r_{0}-4Q^4\right]}}{\sqrt{(Mr_{0}-Q^2)(r_{0}^2-3Mr_{0}+2Q^2)}}
~.\label{margce2}
\end{eqnarray}

\item For Schwarzschild black hole $a=Q=0$, the reciprocal of Critical exponents in terms of ISCO equation  are
\begin{eqnarray}
\frac{1}{\gamma} &=& 2\pi \frac{\sqrt{-[r_{0}-6M]}}{\sqrt{r_{0}-3M}}~.\label{margce3}
\end{eqnarray}

\end{itemize}

\subsection{Critical exponent and Null circular geodesics:}

The reciprocal of Critical Exponent in terms of null circular geodesics is given by

\begin{eqnarray}
\left(\frac{1}{\gamma}\right)_{Null} &=& 2\pi \sqrt{\frac{(L_{c}-aE_{c})^2(3Mr_{c}-4Q^2)(r_{c}^2\mp a\sqrt{Mr_{c}-Q^2})^2}
{r_{c}^6(Mr_{c}-Q^2)}}~.\label{cen1}
\end{eqnarray}

\emph{Special Cases:}

\begin{itemize}

\item For Kerr black hole $Q=0$, the reciprocal of Critical exponent for null circular geodesics are
\begin{eqnarray}
\left(\frac{1}{\gamma}\right)_{Null} &=& 2\pi \sqrt{\frac{3(L_{c}-aE_{c})^2(r_{c}\sqrt{r_{c}}\mp a\sqrt{M})^2}
{r_{c}^5}}~.\label{cen2}
\end{eqnarray}

\item For Reissner Nordstr{\o}m black-hole $a=0$, the  reciprocal of Critical exponent for null circular geodesics are
\begin{eqnarray}
\left(\frac{1}{\gamma}\right)_{Null} &=& 2\pi \sqrt{\frac{L_{c}^2(3Mr_{c}-4Q^2)}{r_{c}^2(Mr_{c}-Q^2)}}~.\label{cen3}
\end{eqnarray}

\item For Schwarzschild black hole $a=Q=0$, and the reciprocal of Critical exponent are
\begin{eqnarray}
\left(\frac{1}{\gamma}\right)_{Null} &=& 2\pi \sqrt{\frac{3L_{c}^2}{r_{c}^2}} ~.\label{cen4}
\end{eqnarray}

\end{itemize}

\section{\label{dis}Discussion}

We have demonstrated that the Lyapunov exponent, KS entropy and Critical exponent can be used to give a full description of  
time-like circular geodesics and null circular geodesics in Kerr Newman  space-time. We then explicitly derived it in terms 
of the radial equation of the ISCO. We proved that the Lyapunov exponent can be used to determine the stability and instability 
of equatorial circular geodesics, both massive and massless particles for Kerr Newman  space-time. The other point we have studied 
that for circular geodesics around the central black-hole, timelike circular 
geodesics  is characterized by the smallest angular frequency as measured by the asymptotic observers-no
other circular geodesics can have a smallest angular frequency. Thus such types of space-times  always have
$\Omega_{timelike}<\Omega_{photon}$ for all time-like circular geodesics. Alternatively it was shown that the 
orbital period of time-like circular geodesics is  characterized by the \emph{longest} orbital
period than the null circular geodesics i.e. $T_{timelike}>T_{photon}$ . This implies  that the timelike circular 
geodesics  provide the \emph{slowest way} to circle the black hole. In fact, any stable timelike circular geodesics other 
than the ISCO traverses more slowly than the null circular geodesics.

\end{document}